\documentclass[doublecol]{epl2} 

\usepackage{amsmath}
\usepackage{graphicx}
\usepackage{dcolumn}
\usepackage{pdfpages}

\title{Barrier-mediated predator-prey dynamics}

\author{Fabian Jan Schwarzendahl \and Hartmut L\"owen}

\institute{                    
  Institut f\"ur Theoretische Physik II: Weiche Materie, Heinrich-Heine-Universit\"at D\"usseldorf, 40225 D\"usseldorf, Germany
}

\abstract{
The survival chance of a prey chased by a predator depends not only on their relative speeds 
but importantly also on the local environment they have to face. 
For example, a wolf chasing a deer might take a long time to cross a river which can quickly be crossed by the deer. 
Here, we propose a simple predator-prey model for a
situation in which both the escaping prey and the chasing predator
have to surmount an energetic barrier.
Different barrier-assisted states of catching or final escaping are
classified and suitable scaling laws separating these two states
are derived. 
We discuss the effect of fluctuations on the catching times and 
determine states in which catching or escaping is more likely. 
We further identify trapping or escaping states which are determined by
hydrodynamics and chemotactic interactions. Our results are of importance for
both microbes and self-propelled unanimate microparticles following each other
by non-reciprocal interactions in inhomogeneous landscapes.}

\begin{document}

\maketitle

\section{Introduction}
The survival chances of animals depend crucially on their ability to
find food and to escape from predators. In the macroscopic world, there is a plethora
of examples where  carnivores follow their prey  trying to catch it but the prey tries to escape:
wolf and deer, lion and wildebeast, shark and fish, etc. Aside from stamina, the
crucial parameter which determines the outcome of a chasing process are the two speeds
$v_1$ and $v_2$ of the prey and the predator. Ideally, on the plane or in three-dimensional space,
when the prey flees straight away from the predator, there will be catching for
$v_1<v_2$ and escaping for $v_1>v_2$. This will be different, however, in an
inhomogeneous landscape where the local speed depends on the details of the environment~\cite{cuddington2002predator,raposo2011landscape,volpe2017topography}.
In particular, an obstacle which will be felt in a different manner by predator and prey
will make the situation more complex such that the simple speed criterion will break down.
Imagine a wolf following a deer which both come close to a river which can be jumped
over by the deer but not by the wolf (the wolf has to slowly swim). Here the obstacle couples
differently to predator and prey and this can decide after all the outcome of the chasing.

While a lot of previous work has modelled predator-prey coupling by coarse-grained density fields~\cite{czirok1996formation,pang2004strategy,tsyganov2003quasisoliton,keller1971model,boraas1998phagotrophy} or by explicit "particles" on a lattice~\cite{kamimura2010group,oshanin2009survival,yang2014aggregation,schwarzl2016single}, agent-based models with
explicit interacting particles which follow each other on continuous
individual trajectories~\cite{sengupta2011chemotactic,angelani2012collective,janosov2017group,surendran2019spatial} were much less considered.
The latter models can particularly be designed
for the mesoscopic world of phagocytes, predatory microbes moving in a fluid or other biological systems~\cite{li2013bifurcation}
in an overdamped way such that inertial effects are absent. Recently there has been a
lot of activity in unanimate predator-prey systems designed
by using synthetic colloidal particles which interact in a non-reciprocal way.
Different realizations involve ion exchange resins building so-called "modular microswimmers"~\cite{ibele2009schooling,niu2017self,niu2018dynamics,niu2018modular,liebchen2018unraveling}, moving droplets following each other~\cite{jin2017chemotaxis,meredith2020predator},
predator-prey-like entities for  active colloidal molecules~\cite{soto2014self,lowen2018active,gonzalez2019directed,schmidt2019light,liebchen2019interactions}, pairs of dust particles in a complex plasma~\cite{ivlev2015statistical,bartnick2016emerging}, and biomimetic active  micromotor
systems~\cite{mou2019active}. Even details of the particle perception can be programmed in synthetic colloidal model systems~\cite{lavergne2019group,bauerle2020formation}. All of these systems naturally experience an inhomogeneous environment (such as confinement, external light intensity etc)
when exhibiting predator-prey characteristics and are thus ideal test cases to study the effect of an energetic barrier on predator-prey dynamics.


In this letter we explore the effect of an energetic barrier on the escape dynamics of a predator-prey system within a simple model of two active particles in the presence of a parabolic potential energy barrier. Both the prey and the predator surmount the barrier but there are different coupling coefficients which make it easier for the prey respective to the predator to overcome the barrier.  Here we propose a one-dimensional model which though simple is general enough to provide an ideal framework to classify different characteristic states for escape and catching in the presence of an obstacle. This model involves overdamped dynamics and is therefore likewise applicable for animate predator-prey system as well as to unanimate self-propelled colloidal pairs with non-reciprocal interactions in case they have to surmount an energetic barrier. The model is partially analytically soluble but flexibly extensible to more complicated couplings such as hydrodynamic interactions and chemotactic sensing.

 We calculate the state diagram of escaping and catching situations in the parameter space and identify scaling laws for the catching time and catching position  at the transition between catching and escaping. Next, fluctuations are included into the motion of both predator and prey. We compute the state diagram and we discuss the effect of noise strength. We then include hydrodynamic and chemotactic couplings between predator and prey and
show their effect on the chasing outcome. We finally discuss the relevance of our results
for both animate and unanimate particles following each other
at low Reynolds number by non-reciprocal interactions.

\section{Ideal predator-prey model}
\begin{figure}[t]
    \centering
    \includegraphics[width=1.0\columnwidth]{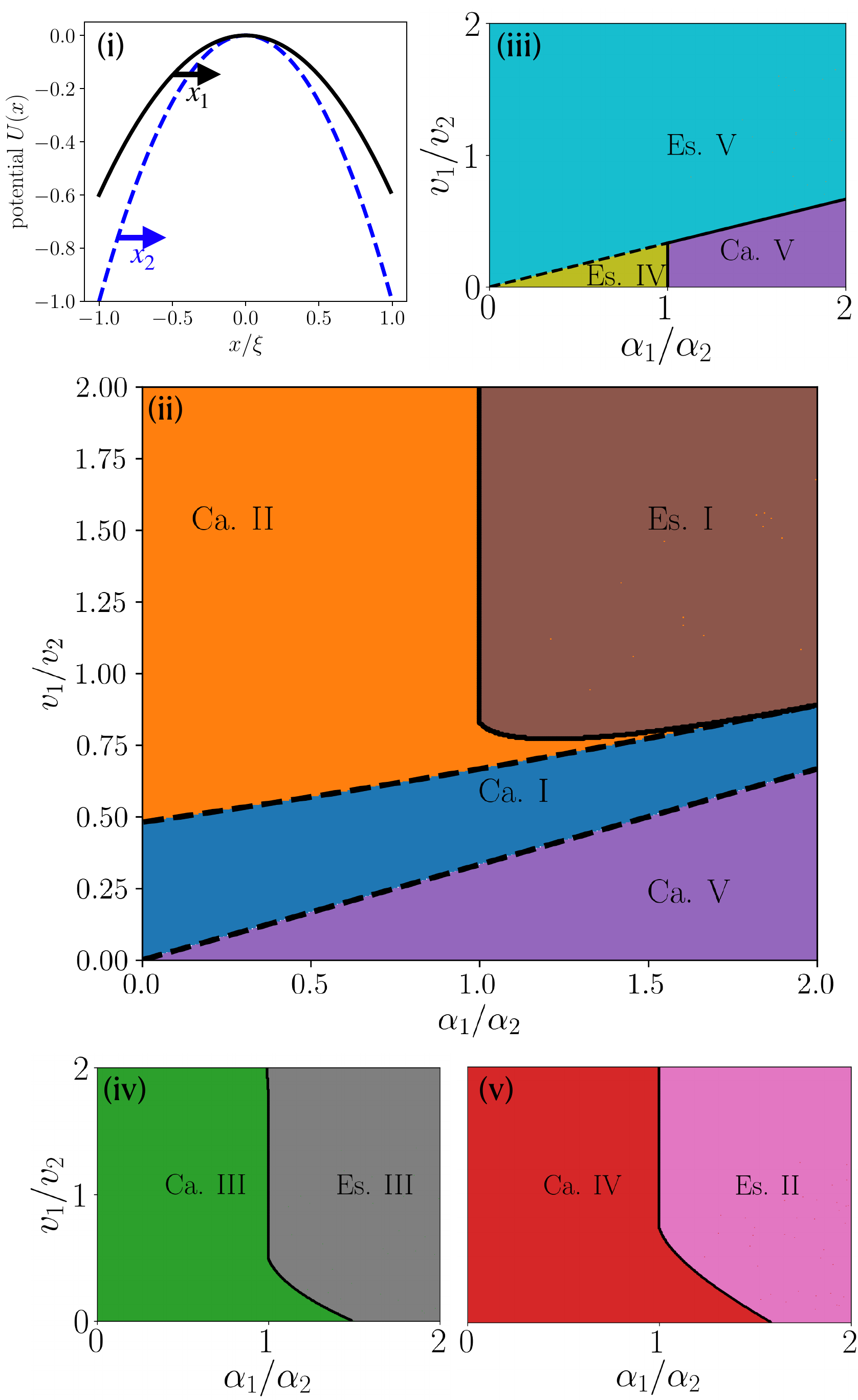}
    \caption{Ideal predator-prey model. (i) Schematic of the prey ($x_1$) and predator ($x_2$) in the presence of a potential barrier $U(x)$ as a function of the reduced one-dimensional coordinate $x/\xi$. The black solid line shows the barrier of the prey and the blue dashed line shows the barrier of the predator. (ii) State diagram with initial conditions $x_1(0)=-\xi/3$ and $x_2(0)=-\xi/2$ showing catching and escaping regions for varying $v_1/v_2$ and $\alpha_1/\alpha_2$. For classification see Table~\ref{tab:cases}. (iii)-(v): State diagrams for different initial conditions. ((iii): $x_1(0)=-\xi/3$, $x_2(0)=-3\xi/2$; (iv): $x_1(0)=\xi/4$, $x_2(0)=-\xi/4$; (v): $x_1(0)=\xi/2$, $x_2(0)=\xi/4$) }
    \label{fig:simple_model}
\end{figure}
We consider a one dimensional model of predator and prey which are crossing a potential barrier. Figure~\ref{fig:simple_model}(i) shows a schematic of prey $x_1(t)$ and predator $x_2(t)$ in the presence of their respective potential barrier $U(x_{1,2})$, where both are moving into the positive $x$-direction. Since we are motivated by microswimmers we are working in the low Reynolds number limit and assume the motion of predator and prey to be overdamped. The equations of motion for the position of the prey and the position of the predator are given by
\begin{align}
    \dot{x}_1=v_1+\alpha_1x_1, \label{eq:simple_x1}
\\
    \dot{x}_2=v_2+\alpha_2x_2,
    \label{eq:simple_x2}
\end{align}
where $v_1,v_2$ are the self-propulsion speeds, and $\alpha_1,\alpha_2$ are coupling constants to the respective potential barrier. 
Equations~\eqref{eq:simple_x1}-\eqref{eq:simple_x2} have the solutions
\begin{align}
x_1(t)=\frac{1}{\alpha_1}((v_1 +\alpha_1 x_1(0))e^{\alpha_1 t} - v_1),
 \label{eq:sol_simple_x1}
\\
x_2(t)=\frac{1}{\alpha_2}((v_2 +\alpha_2 x_2(0))e^{\alpha_2 t} - v_2),
 \label{eq:sol_simple_x2}
\end{align}
with initial conditions $x_{1,2}(0)$. 
We use $\tau=1/\alpha_2$ as a natural unit of time and $\xi= v_2/\alpha_2$ as a natural length scale, which are the physical time and length scales related to the predator.

\begin{table*}[!ht ]
    \centering
    \begin{tabular}{|c|c|c|c|}
    \hline
        case  & description & initial conditions & catching condition
    \\ \hline
        Ca.~I  & caught while summiting the barrier & $x_1(0)<0$, $x_2(0)<x_1(0)$ & $x^*>x_1(0)$ 
    \\ \hline
        Ca.~II  & caught after summiting the barrier  & $x_1(0)<0$, $x_2(0)<x_1(0)$ & $x^*>0$ 
    \\ \hline
        Ca.~III  &caught after prey summits  & $x_1(0)>0$, $x_2(0)\leq0$ & $x^*>0$ 
    \\ \hline
        Ca.~IV  & caught descending barrier  & $x_1(0)>0$, $x_2(0)\geq0$ & $x^*>0$
    \\ \hline        
        Ca.~V  & caught descending barrier & $x_1(0)<0$, $x_2(0)<x_1(0)$ & $x^*<x_1(0)$ \\& without summiting & & 
    \\ \hline   
        Es.~I  & both are summiting the barrier  & $x_1(0)<0$, $x_2(0)<x_1(0)$ & $x_1(\infty)=x_2(\infty)=\infty$
    \\ \hline 
        Es.~II  & both descending in positive direction & $x_1(0)>0$, $0<x_2(0)<x_1(0)$ & $x_1(\infty)=x_2(\infty)=\infty$
    \\ \hline 
        Es.~III  & both descending in opposite directions  & $x_1(0)>0$, $x_2(0)<0$ & $x_1(\infty)=\infty$, $x_2(\infty)=-\infty$
    \\ \hline        
        Es.~IV  & both descending in negative direction  & $x_1(0)<0$, $x_2(0)<x_1(0)$ & $x_1(\infty)=x_2(\infty)=-\infty$
    \\ \hline
            Es.~V  & only prey is  summiting the barrier  & $x_1(0)<0$, $x_2(0)<0$ & $x_1(\infty)=\infty$, $x_2(\infty)=-\infty$
    \\ \hline 
    \end{tabular}
    \caption{
    Classification of catch (Ca.) and escape (Es.) cases, where $x_1$ is the prey and $x_2$ is the predator.}
    \label{tab:cases}
\end{table*}

Single active particles similar to our Eq.~\eqref{eq:simple_x1} which are crossing a barrier have been studied theoretically in one dimensional landscapes~\cite{caprini2019active,geiseler2016kramers,sharma2017escape,dhar2019run,blossey2019chromatin}. 
An experimentally realizeable system that is expected to have similar dynamics to our Eq.~\eqref{eq:simple_x1}-\eqref{eq:simple_x2} consists of two self-propelled droplets that chase each other such as in \cite{meredith2020predator}, and are confined into a one dimensional microfluidic channel~\cite{jin2017chemotaxis,deblois2021swimming}. Additionally, the microfluidic channel has a physical barrier, that the droplets have to overcome, this physical barrier acts as potential barrier by means of the gravitational force. Here, the self propulsion velocities can be tuned by the chemical compositions of the surrounding medium~\cite{maass2016swimming} and the coupling constants of the potential barrier can be tuned via the droplets' size. 
A barrier can also be realized by viscosity gradients in a surrounding fluid medium or external flow fields created in a microfluidic device~\cite{liebchen2019optimal}.

In the following we want to distinguish the scenarios in which the predator can catch the prey and where it can not. In order to determine catching we use the catching time $t^*$ given by the condition
\begin{align}
    x_1(t^*)=x_2(t^*).
    \label{eq:catching_condition}
\end{align}
Additionally, we use the catching position $x^*=x_{1,2}(t^*)$, which shows where the prey is caught. By considering the initial conditions and long time limits of Eq.~\eqref{eq:sol_simple_x1}-\eqref{eq:sol_simple_x2} we can categorise five different catching cases and five different escaping cases which are summarized in Table \ref{tab:cases}.

Figure~\ref{fig:simple_model}(ii) shows the catching and escaping states for varying $\alpha_1/\alpha_2$ and $v_1/v_2$ with initial conditions $x_1(0)=-\xi/3<0$ and $x_2(0)=-\xi/2<0$ (here the condition Eq.~\ref{eq:catching_condition} was solved numerically). We find three different catching states, where in Ca.~I the prey is caught while summiting the barrier, in Ca.~II  the prey is caught after summiting the barrier and for Ca.~V the prey is caught descending barrier without summiting it. Furthermore, we find one escaping region (Es.~I), where both are summiting the barrier.

We continue by analyzing the lines dividing the respective regions in Fig.~\ref{fig:simple_model}(ii). The line separating the region Ca.~I and Ca.~II is determined by the fact that 
catching happens on top of the barrier, meaning that $x^*=0$. 
In region Ca.~I the prey can cross the barrier, while in Ca.~V it can not. Therefore, the line separating regions Ca.~I and Ca.~V  can be determined from  the longtime limits which gives $\frac{v_1}{v_2}=-x_1(0) \frac{\alpha_1}{\alpha_2}$. 

The transition from the escape region Es.~I and the catching regions was determined numerically. When we approach Es.~I from below while increasing $v_1/v_2$, we find that at the transition line the catching time stays finite.
For large $\alpha_1/\alpha_2$ this can be rationalized since we are going from Ca.~I to Es.~I. Here, the dividing line between catching and escaping approaches the line at which $x^*=0$, meaning that catching happens before the barrier, or the prey escapes. Hence, the catching time stays finite since  $t(x^*=0)$ is finite. 
This can also be seen when we solve Eq.\eqref{eq:catching_condition} for the special case $\alpha_1=2\alpha_2=2\alpha$ which stays finite (see SI).

On the other hand as we approach the escape region from the left (from Ca.~II), we find that the catching time and position both diverge. To obtain an understanding of the scaling of the divergence, we approximate our solutions (Eq.\eqref{eq:simple_x1}-\eqref{eq:simple_x2}) for barrier dominated motion (see SI). We find that as we approach $\alpha_1 \rightarrow \alpha_2$, the catching time scales as $ t^*\sim 1/(\alpha_1-\alpha_2)$. Intuitively, for $\alpha_1 > \alpha_2$ the self propulsion velocity of the predator is not sufficient anymore to catch the prey since the motion of both predator and prey is dominated by them descending the barrier. Similarly, as we come closer to from Ca.~II to Es.~I the catching dynamics becomes dominated by the potential barrier and the importance of the self-propulsion decreases.
Here, it is interesting to see what happens at $\alpha_1=\alpha_2=\alpha$ when it is approached from below. This case can be solved exactly (see SI) where we find that $t^*$ diverges as $\alpha \rightarrow \frac{v_2-v_1}{x_1(0)-x_2(0)}$. Going along the line dividing the regions Ca.~II and Es.~I the point $\alpha_1=\alpha_2=\alpha$ is where the catching time starts to diverge and is thus consistent with the previous analysis.

Furthermore, we analysed the scaling of the relative distance, which to first order reads
\begin{align}
    x_1(t)-x_2(t) \approx A (t-t^*) + ...
    \label{eq:xoft_scaling}
\end{align}
The prefactor $A$ scales in the limit $\alpha_1 \rightarrow \alpha_2$ as $\mathrm{ln} A \sim 1/(\alpha_1-\alpha_2)$ (see SI for a details). 
Similar to the catching time, the relative position of predator and prey diverges, since their dynamics is dominated by them descending the potential barrier. 

We continue by analyzing different initial conditions, that lead to other catching and escape scenarios. Figure~\ref{fig:simple_model}(iii) shows the  state diagram for $x_1(0)=-\xi/3$ and $x_2(0)=-3\xi/2$, where we find the catching case Ca.~V in which we have catching while both are descending the barrier without summiting and the escape cases Es.~IV, where both descend into the negative direction, as well as the Es.~V case where only the prey is able to summit the barrier.  Here, the dividing lines between all respective regions were determined from the long time limits of solutions Eq.\eqref{eq:sol_simple_x1}-\eqref{eq:sol_simple_x2}.

Figure~\ref{fig:simple_model}(iv) has initial conditions $x_1(0)=\xi/4$ and $x_2(0)=-\xi/4$ where we have one catching case Ca.~III in which the prey is caught after summiting and escaping case Es.~III where predator and prey descend into opposite directions. Here, the dividing line between  Ca.~III and Es.~III was determined numerically, however, the scaling arguments described above are  still valid. Similarly, Fig.~\ref{fig:simple_model}(v) with initial condition $x_1(0)=\xi/2$ and $x_2(0)=\xi/4$ has one catching case Ca.~IV where the catching happens while descending the barrier and one escape scenario Es.~II where both are descending in the positive direction and the above scaling arguments still hold.

In the following we will extend our ideal predator-prey model (Eq.~\eqref{eq:simple_x1}-\eqref{eq:simple_x2}) to account for fluctuations, chemotactic and hydrodynamic interactions. Here, we will restrain ourselves to the initial conditions $x_1(0)=-\xi/3$ and $x_2(0)=-\xi/2$, since the essential phenomena of our model are captured within these conditions.

\section{Predator-prey model with fluctuations}
\begin{figure*}[ht]
    \centering
    \includegraphics[width=1.0\textwidth]{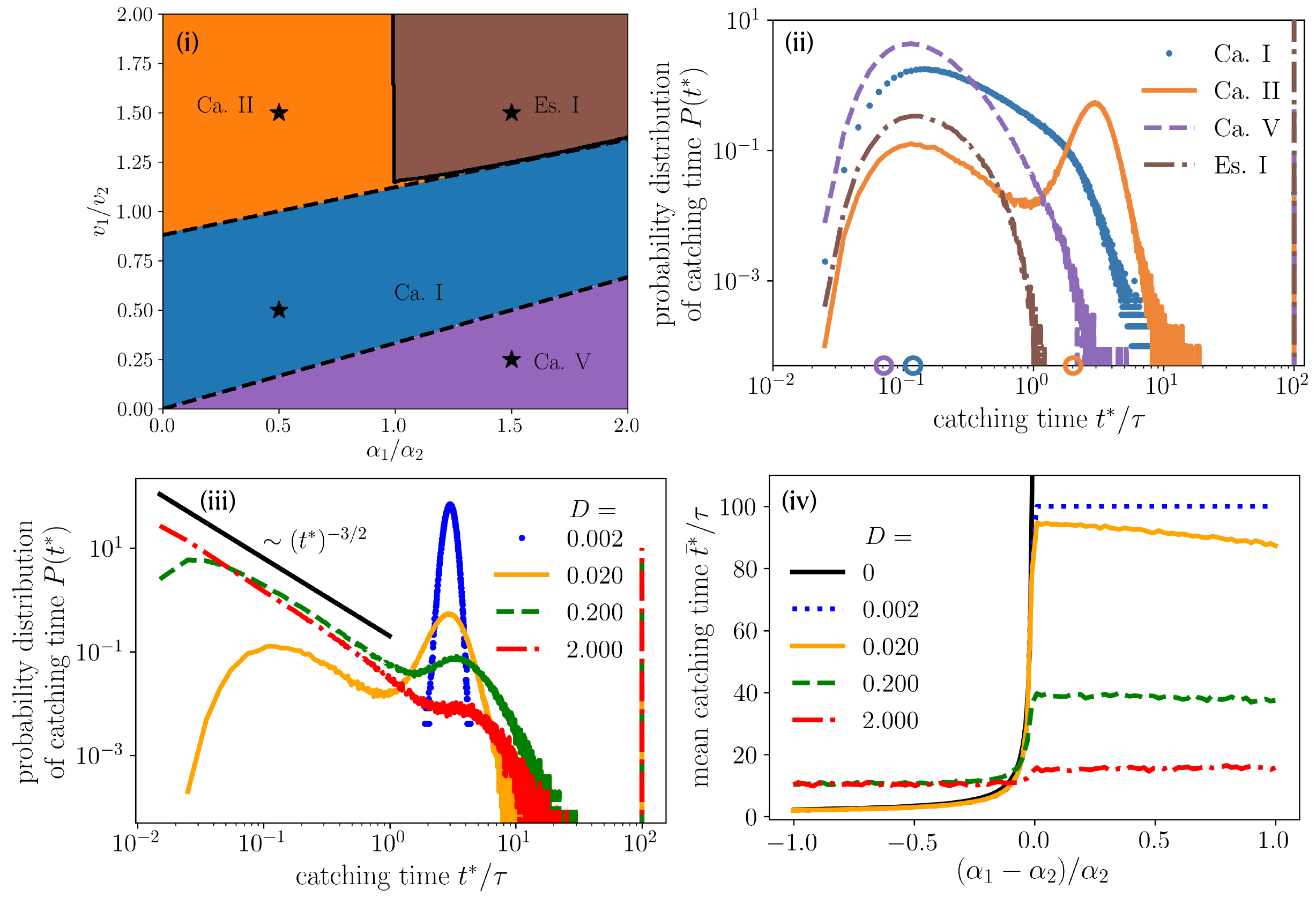}
    \caption{Predator-prey model with fluctuations. (i): State diagram for different catching and escape cases for varying $\alpha_1/\alpha_2$ and $v_1/v_2$ (fixed $D=0.02 \xi^2/\tau$). The classification can be found in Table~\ref{tab:cases}. Stars show the values used for the representative catching time distributions in (ii). (ii): Representative catching time distribution for one example of each catching and escape case found in (i) (fixed $D=0.02 \xi^2/\tau$). Circles on the catching time axis show the catching condition Eq.~\eqref{eq:diff_crit} (iii): Catching time distribution for different values of noise strength $D$. Black solid line shows a $P(t^*)\sim(t^*)^{-3/2}$ scaling. (iv): Mean catching time as function of the relative coupling parameters for different noise strength $D$.  }
    \label{fig:diffusive_model}
\end{figure*}

We now extend our predator-prey model to account for fluctuations of both predator and prey. Our equations of motion are 
\begin{align}
    \dot{x}_1=v_1+\alpha_1x_1 + f_1, \label{eq:diffusion_x1}
\\
    \dot{x}_2=v_2+\alpha_2x_2 + f_2,
    \label{eq:diffusion_x2}
\end{align}
where $f_1$ and $f_2$ are Gaussian random forces with $\langle f_i(t) \rangle =0$ and $\langle f_i(t) f_i(t') \rangle =2 D \delta(t-t')$. Here, $D$ is the noise strength and $\delta(*)$ is the Dirac-delta function. 
The random forces introduced here can stem from fluctuations of a surrounding fluid, however, they do not need to obey a fluctuation dissipation theorem since they can also be introduced by biological fluctuations (in case of biological predator and prey).
For fluctuating predator and prey the catching time and position are now smeared out by the random forces, $f_1$ and $f_2$ such that we need a new definition of the catching criterion. We assume a "worst case" for the prey in which we reduce the mean position of the prey by its variance and enhance the position of the predator by its variance, which means
\begin{align}
    \bar x_1(t^*) -\sqrt{ \Delta_1(t^*)} =\bar x_2(t^*) +\sqrt{ \Delta_2(t^*)},
    \label{eq:diff_crit}
\end{align}
where $\bar x_i(t)= \langle x_i(t) \rangle$ is the mean value of $x_i(t)$ and $\Delta_i(t) =  \langle (x_i(t)  - \bar x_i(t) )^2\rangle$ is the mean-square-displacement. The mean value is the solution in the absence of fluctuations (Eq.~\eqref{eq:sol_simple_x1}-\eqref{eq:sol_simple_x2}) and the mean-square-displacement is $\Delta_i(t) =  \frac{D}{\alpha_i} (e^{2 \alpha_i t} -1) $. 
Using the criterion Eq.~\eqref{eq:diff_crit} we find the state diagram shown in Fig.~\ref{fig:diffusive_model}(i) where we used a fixed noise strength. Similar to the situation without noise (Fig.~\ref{fig:simple_model}(ii)), we find three catching and one escape scenario, however, the relative size of the regions is changed by noise. Here, the fluctuations can help the predator to catch the prey.

To further investigate the effect of fluctuations we numerically solved Eq.~\eqref{eq:diffusion_x1}-\eqref{eq:diffusion_x2} and extracted the catching time distributions shown in Fig.~\ref{fig:diffusive_model}(ii) where we show a distribution for each catching or escaping case found in Fig.~\ref{fig:diffusive_model}(i).  
For catching case Ca.~I we find a broad distribution that has its maximum at $t^*/\tau \approx 0.1$ and then exhibits a large shoulder towards higher catching times. In case of Ca.~II we find a bimodal distribution, with a maximum at $t^*/\tau \approx 3$ stemming from the deterministic dynamics and at $t^*/\tau \approx 0.1$ induced  by fluctuations, which means that the prey is caught before summiting. In case of Ca.~V the distribution only has a single maximum and is centered around $t^*/\tau \approx 0.1$. For Es.~I we find that fluctuations can cause catching for early times, however, we find a large peak at $t^*/\tau \approx 100$, which corresponds to escaping as this is our maximal simulation time. Note, that for all four cases we find a peak at $t^*/\tau \approx 100$, which should be categorised as escaping. For the catching cases Ca.~I, Ca.~II and Ca.~V this is a "lucky" fluctuation-induced escaping of the prey.

Next, we test the dependence of our results on the strength of the noise $D$. Figure~\ref{fig:diffusive_model}(iii) shows the catching time distribution for the case Ca.~I ($\alpha_1/\alpha_2=0.5$, $v_1/v_2=1.5$) at different noise strength. For small noise strength ($D=0.002 \xi^2/\tau$) we find a sharp peak, and here fluctuations have minor effects. Going to higher values ($D=0.02 \xi^2/\tau$) the distribution becomes bimodal and then ($D=0.2 \xi^2/\tau$, $D=2 \xi^2/\tau$) spreads out to very low catching time values, with an approximate scaling $P(t^*) \sim (t^*)^{-3/2}$. For the latter, the catching process is dominated by fluctuations and the problem reduces to finding the first hitting time of a one dimensional Brownian particle, which has the known scaling with an exponent of $-3/2$~\cite{redner2001guide}, consistent with out finding. Again, for very large times ($t^*/\tau \approx 100$), all probability distributions show a peak, which signals escaping by fluctuations.

Continuing, we study the mean catching time for varying coupling ratio in Fig.~\ref{fig:diffusive_model}(iv). In the small noise limit, ($D=0.002 \xi^2/\tau$) the catching time diverges as we approach $\alpha_1 \rightarrow \alpha_2$, as also seen in the deterministic case ($D=0$) (note that the plateau value for $D=0.002 \xi^2/\tau$ corresponds to the maximal simulation time). Going to higher noise strength the catching time is still enhanced for $\alpha_1 \rightarrow \alpha_2$, however, we find that the plateau value of mean catching time for $\alpha_1>\alpha_2$ is decreased, representing the fact that fluctuations can lead to catching. Interestingly, for  $D=0.02 \xi^2/\tau$ we find a small decay of the plateau value of the catching time for $(\alpha_1-\alpha_2)/\alpha_2 \rightarrow 1$. This is due to the fact that for larger $\alpha_1$ the prey needs a longer time to overcome the barrier and thus there is an enhancement of catching due to noise before crossing the barrier.

\begin{figure*}[ht]
    \centering
    \includegraphics[width=1.0\textwidth]{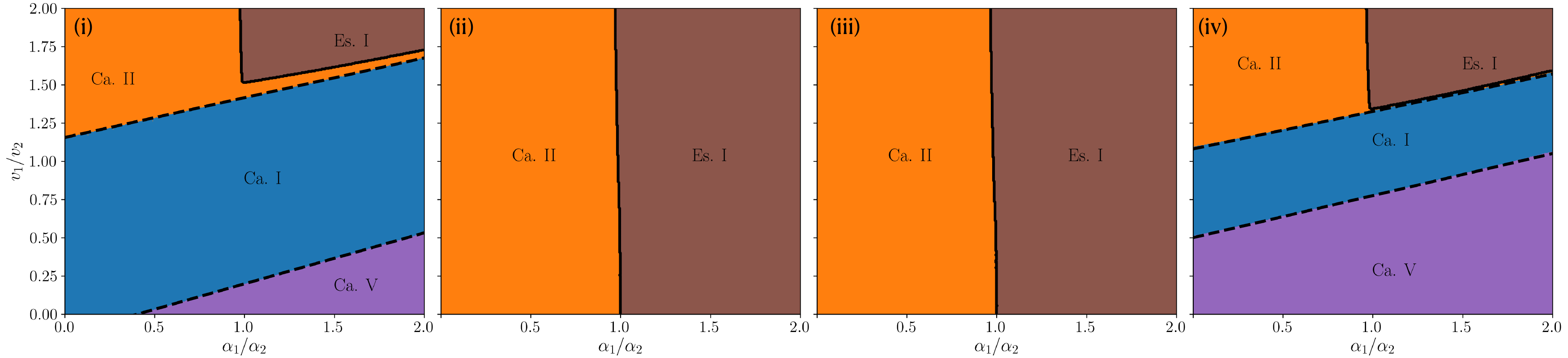}
    \caption{
    State diagrams for models including chemotaxis and hydrodynamic interactions for varying $v_1/v_2$ and $\alpha_1/\alpha_2$. (i): chemotaxis. (ii): force monopole. (iii): pusher-type swimmer (iv): puller-type swimmer. State classification is according to Table~\ref{tab:cases}. }
    \label{fig:chemo_hydro_model}
\end{figure*}

\section{Chemotactic and hydrodynamic interactions}
To make a connection to microswimmers we study our predator-prey system in the presence of chemotactic and hydrodynamic interactions. Artificial droplet swimmers such as in~\cite{meredith2020predator,jin2017chemotaxis,deblois2021swimming} interact via chemotaxis and hydrodynamic interactions have been shown to play an important role in suspensions of microswimmers~\cite{schwarzendahl2019hydrodynamic,zottl2016emergent,theers2018clustering,maass2016swimming,bechinger2016active,elgeti2015physics}.

We consider the situation where predator and prey interact through a chemical field. Both secrete a chemical which induces a force on the respective other swimmer (see SI for details, \cite{sengupta2011chemotactic,pohl2014dynamic,soto2014self}). We assume that the forces are proportional to the gradient of the chemical field which quickly relaxes to its stationary distribution. This gives rise to the following equations of motion
\begin{align}
    \dot{x}_1=v_1+\alpha_1x_1 + A_1 \frac{1}{(x_1-x_2)^2}, \label{eq:chemo_x1}
    \\
    \dot{x}_2=v_2+\alpha_2x_2+ A_2 \frac{1}{(x_1-x_2)^2},
    \label{eq:chemo_x2}
\end{align}
where $A_1$ and $A_2$ control the strength of chemoattraction or chemorepulsion.

Figure~\ref{fig:chemo_hydro_model}(i) shows the resulting state diagram for catching and escaping, where we used $A_1=0.001\xi^3/\tau$ and $A_2=0.02\xi^3/\tau$. The catching and escaping regions that we find are the same as for the ideal model (see Fig.~\ref{fig:simple_model}(ii)), however, the relative sizes of the regions is changed by chemotactic interactions. Specifically, chemotaxis enhances the Ca.~I case and reduces the extend of the escaping region Es.~I, due to effective attraction between predator and prey. 

Next, we consider the effect of a fluid surrounding predator and prey leading to hydrodynamic interactions. 
First, we investigate the situation in which both predator and prey act as a force monopole, leading us to the equations
\begin{align}
    \dot{x}_1=&v_1+\alpha_1x_1+ \frac{\beta}{4\pi\eta}\left(\alpha_2\mathrm{sign}(x_1-x_2)+ \frac{v_2}{|x_1-x_2|}\right), \label{eq:monopole_x1}
\\
    \dot{x}_2=&v_2+\alpha_2x_2 + \frac{\beta}{4\pi\eta}\left(\alpha_1  \mathrm{sign}(x_2-x_1)+ \frac{v_1}{|x_1-x_2|}\right),
    \label{eq:monopole_x2}
\end{align}
where $\eta$ is the fluids' viscosity, $\beta$ is a parameter that depends on the geometric details of the swimmer and sign$(*)$ is the sign function (for a derivation of the interactions see SI).

In Fig.~\ref{fig:chemo_hydro_model}(ii) we show the resulting state diagram, where we only find two cases Ca.~II and Es.~I (here $\beta=1/\tau$,$\eta=1/(\tau\xi)$). The hydrodynamic interaction between predator and prey leads to an effective repulsion, such that it is easier for the prey  to escape, enhancing the Es.~I region. Similarly, the Ca.~I and Ca.~V region vanish, since the predator and prey are repelled, making it necessary to first cross the border in order for predator and prey to come close enough for catching.

In a second step predator and prey induce a hydrodynamic flow field corresponding to a force dipole. Here, we use the equations
\begin{align}
    \dot{x}_1=&v_1+\alpha_1x_1 + \frac{\beta \mathrm{sign}(x_1-x_2)}{4\pi\eta}\left(\alpha_2+ \frac{v_2 \lambda}{(x_1-x_2)^2}\right), \label{eq:dipole_x1}
\\
    \dot{x}_2=&v_2+\alpha_2x_2  + \frac{\beta \mathrm{sign}(x_2-x_1)}{4\pi\eta}\left(\alpha_1+ \frac{v_1\lambda}{(x_1-x_2)^2}\right),
    \label{eq:dipole_x2}
\end{align}
where the sign of $\lambda$ decides whether we have puller- ($\lambda<0$) or pusher-type ($\lambda>0$) swimmers (for a derivation see SI). 

The state diagram for pusher-type swimmers ($\lambda=0.1\xi$, $\beta=1/\tau$, $\eta=1/(\tau\xi)$) is shown in Fig.~\ref{fig:chemo_hydro_model}(iii). Here, the situation is similar to the force monopole. We find one catching region (Ca.~II) and one escaping region (Es.~I). The pusher-type hydrodynamic interactions introduce an effective repulsion between predator and prey, which gives rise to larger catching times and subsequently the  enhancement of the Ca.~II region. Similarly, the repulsion leads to the fact that catching happens less often and thus an increase of the Es.~I region. 

For puller type swimmers ($\lambda=-0.1\xi$, $\beta=1/\tau$, $\eta=1/(\tau\xi)$) we find the state diagram shown in Fig.~\ref{fig:chemo_hydro_model}(iv), which has three catching and one escape regions, similar to the ideal case (Fig.~\ref{fig:simple_model}(ii)). Here, the catching regions are enhanced since the puller-type hydrodynamic interactions give an effective attraction between predator and prey~\cite{guzman2016fission}. Interestingly, the Ca.~V region is larger than in the ideal case, which shows that the attraction between predator and prey enhances catching before the barrier.

\section{Conclusions}
In conclusion, we have introduced a one dimensional predator-prey model in the presence of a potential barrier. We classified different catching and escaping states, calculated state diagrams displaying the occurrence of these states and determined scaling laws. We extended our model to account for fluctuations, computed a state diagram and showed that it qualitatively agrees with our ideal model. Here, the relative size of the catching regions is increased, since catching can be induced by fluctuations. We varied the noise strength and discussed the effect on catching times. Furthermore, we included chemotactic and hydrodynamic interactions. Chemotactic and puller-type swimmer interaction give a qualitatively similar state diagram as our ideal model, with an enhancement on catching states due to effective attractive interactions. On the other hand, pusher-type swimmers and hydrodynamic monopoles decrease the catching regions and enhance escaping, due to effective repulsion between predator and prey. 

Our model makes testable predictions about the outcomes of predator-prey dynamics in the presence of a potential barrier in one dimension. A possible experimental realization of our model consists of two active droplets chasing each other~\cite{meredith2020predator} and  encountering a physical barrier. Other realizations of barriers could be achieved by flow fields using microfluidic devices or by means of viscosity gradients. Furthermore, our model is relevant to microbial systems with predator-prey dynamics, that stem from chemotactic interactions. 

In future work we aim to extend our model to two dimensional landscapes~\cite{liebchen2019optimal,yang2018optimal,zimmermann2021negative}, which might be realized by viscosity gradients~\cite{coppola2021green,datt2019active,dandekar2020swimming} or external flow fields~\cite{berman2021transport} and account for more realistic microswimmer models. Also, we will include inertial effects \cite{scholz2018inertial,lowen2020inertial,sprenger2021time,caprini2021inertial} to make a connection with macroscopic predator prey dynamics.


\acknowledgments
We acknowledge funding by the DFG grand LO 418/23.
\bibliographystyle{eplbib}
\bibliography{predatorprey}

\clearpage


\section*{Supplementary material}

# Supplementary Information: Barrier-mediated predator-prey dynamics

Fabian Jan Schwarzendahl and Hartmut Löwen

*Institut für Theoretische Physik II: Weiche Materie, Heinrich-Heine-Universität Düsseldorf, 40225 Düsseldorf, Germany*

March 22, 2021

## 1 Phase flows

### 1.1 Ideal model

For each respective region we show a phase flow in Fig. S1(ii)-(v), where the green circle shows the initial condition used in Fig. S1(i) and the black solid line shows the catching condition Eq. (5)(main text). Consistently, for the catching cases Ca. I, Ca. II, and Ca. V the flow lines starting from the initial condition point towards the catching condition, showing catching. On the other hand for Es. I the flow lines starting from the initial condition point away from the catching condition, showing escaping.

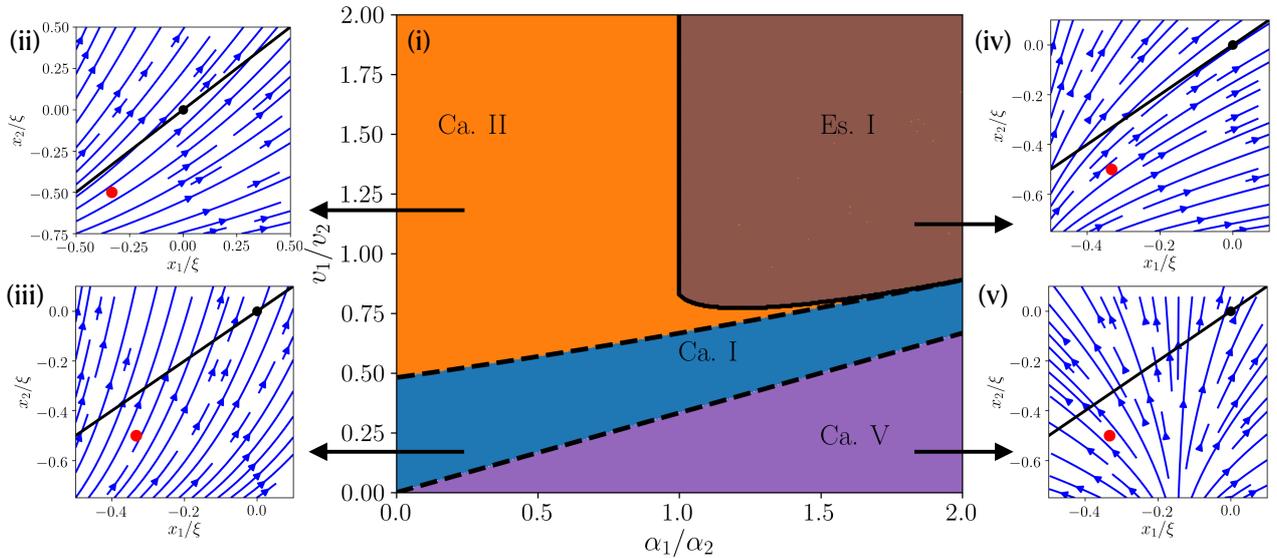

Figure S1: Ideal predator-prey model. (i) State diagram with initial conditions $x_1(0) = -\xi/3$ and $x_2(0) = -\xi/2$ showing catching and escaping regions for varying $v_1/v_2$ and $\alpha_1/\alpha_2$. For classification see Table 1 (main text). (ii)-(v): Representative phase flows for Ca. II (ii) ($\alpha_1/\alpha_2 = 0.5$, $v_1/v_2 = 0.75$), Ca. I (iii) ($\alpha_1/\alpha_2 = 0.25$, $v_1/v_2 = 0.25$), Es. I (iv) ($\alpha_1/\alpha_2 = 1.7$, $v_1/v_2 = 1$) and Ca. V (v) ($\alpha_1/\alpha_2 = 1.7$, $v_1/v_2 = 0.25$). Blue lines show the phase flow, red circles the initial conditions used in (i), black circles the origin and black line the catching condition Eq. (5) (main text).

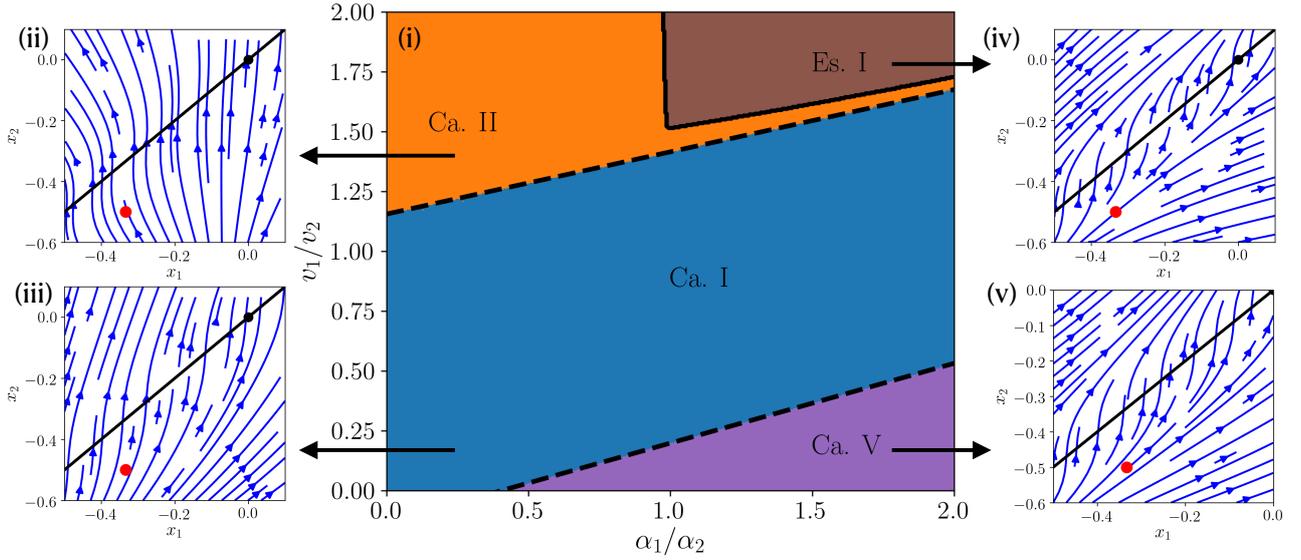

Figure S2: Chemotactic model. (i): State diagram showing the different catching and escaping cases that we find for varying $\alpha_1/\alpha_2$ and $v_1/v_2$ with initial conditions $x_1(0) = -\xi/3$ and $x_2(0) = -\xi/2$. (ii)-(v): Phase flows for the different cases Ca. II (ii) ($\alpha_1/\alpha_2 = 1.5$, $v_1/v_2 = 0.1$), Ca. I (iii) ($\alpha_1/\alpha_2 = 0.5$, $v_1/v_2 = 0.5$), Es. I (iv) ($\alpha_1/\alpha_2 = 1.75$, $v_1/v_2 = 1.75$) and Ca. V (v) ($\alpha_1/\alpha_2 = 0.1$, $v_1/v_2 = 1.1$). The blue lines with arrows show the phase flow, the red circle the initial conditions used in (i), the black circle the zero point and the black line the catching condition Eq. (5) (main text).

### 1.2 Chemotatic model

We show the phase flow for each respective catching and escaping region in Fig. S2(ii)-(v), where the red circle shows the initial condition used in Fig. S2(i). For the catching cases (Fig. S2(ii),(iii),(v)) the flow lines starting from the initial condition lead towards the black line, which represents the catching condition Eq. (5)(main text). On the other hand, for the escaping case Es. I (Fig. S2(iv)) the flow lines lead away from the catching condition.

## 2 Scaling law calculations

The dividing line between Ca. I and Ca. II in Fig. 2(ii) (main text) is determined by the condition $x^* = 0$, which yields

$$\frac{v_1}{v_2} = -\frac{\alpha_1 x_1(0)(1 + \alpha_2 x_2(0))^{-\alpha_1/\alpha_2}}{-1 + (1 + \alpha_2 x_2(0))^{-\alpha_1/\alpha_2}}. \tag{S1}$$

As explained in the main text, at the transition from Ca. I to Es. I the catching time stays finite. Here, we can solve Eq. 5 (main text) for the special case $\alpha_1 = 2\alpha_2 = 2\alpha$, where we find the catching time

$$\begin{aligned}t^* = &\frac{1}{\alpha}\ln[(v_2/\alpha + x_2(0) \\ &- [v_1^2 - 2v_1v_2 + v_2^2 + 2\alpha v_1 x_1(0) - 4\alpha v_2 x_1(0) \\ &+ 2\alpha v_2 x_2(0) + \alpha^2 x_2(0)^2]^{1/2}/\alpha)/(v_1/\alpha + 2x_1(0))],\end{aligned} \tag{S2}$$

which does not diverge.

The catching time and position both diverge as we approach the escape region from the left (from Ca. II).

In order to make progress analytically for this limit, we approximate the solutions in Eq. 1-2 (main text) by

$$x_i(t) = \frac{1}{\alpha_i}((v_i + \alpha_i x_i(0))e^{\alpha_i t}). \tag{S3}$$

This approximation is justified since the dynamics is dominated by the particles going down the potential barrier (here $i = 1, 2$). Within this approximation we can analytically determine the following catching time

$$t^* = \frac{1}{k(\alpha_1 - \alpha_2)} \ln\left[\frac{\alpha_1(v_2 + \alpha_2 x_2(0))}{\alpha_2(v_1 + \alpha_1 x_1(0))}\right]. \tag{S4}$$

This results in the scaling $t^* \sim 1/(\alpha_1 - \alpha_2)$ for $\alpha_1 \to \alpha_2$.

For the special case $\alpha_1 = \alpha_2 = \alpha$ the catching time can be found exactly

$$t^* = \frac{1}{\alpha} \ln\left(\frac{v_1 - v_2}{v_1 - v_2 + \alpha x_1(0) - \alpha x_2(0)}\right), \tag{S5}$$

which diverges for $\alpha \to \frac{v_2 - v_1}{x_1(0) - x_2(0)}$.

By analysing the scaling of the relative distance

$$x_1(t) - x_2(t) \approx A(t - t^*) + ..., \tag{S6}$$

we computed the parameter $A$, that can be found using the approximate catching time Eq.(S4) giving

$$A = (v_1 + \alpha_1 x_1(0))\left[\frac{\alpha_1(v_2 + \alpha_2 x_2(0))}{\alpha_2(v_1 + \alpha_1 x_1(0))}\right]^{\frac{\alpha_1}{\alpha_1 - \alpha_2}}$$
$$- (v_2 + \alpha_2 x_2(0))\left[\frac{\alpha_1(v_2 + \alpha_2 x_2(0))}{\alpha_2(v_1 + \alpha_1 x_1(0))}\right]^{\frac{\alpha_2}{\alpha_1 - \alpha_2}}, \tag{S7}$$

which in the limit $\alpha_1 \to \alpha_2$ scales as $\ln A \sim 1/(\alpha_1 - \alpha_2)$.

## 3 Chemotatic interactions

We model the chemotactic interaction between predator and prey through a chemical field $c(x, t)$ which evolves according to

$$\partial_t c(x, t) = D_c \partial_x^2 c(x, t) + \sum_i \gamma_i \delta(x - x_i), \tag{S8}$$

where $D_c$ is the diffusion constant of the chemical, $\gamma_i$ is the amount of chemical added by predator or prey (here $i = 1, 2$), and $\delta(.)$ is the Dirac delta function. We now assume that the relaxation of the chemical is quick which means that we can assume a stationary distribution $c(x, t) = c(x)$. The resulting chemical field is

$$c(x) = \frac{1}{4\pi D_c} \sum_i \gamma_i \frac{1}{|x - x_i|}. \tag{S9}$$

Furthermore, we assume that each particle only reacts to the chemical secreted by the other particle and the resulting force $f_{i,\text{chem}}$ is proportional to the gradient of the chemical field giving

$$f_{i,\text{chem}} = A_i \frac{1}{(x_1 - x_2)^2}. \tag{S10}$$

3

# 4 Hydrodynamic interactions

We consider that our predator and prey move in a three dimensional low Reynolds number fluid, such that the fluids' dynamics is governed by the Stokes equations

$$\eta \nabla^2 \boldsymbol{u} = \nabla p - \boldsymbol{f}^{\text{ext}}, \quad \nabla \cdot \boldsymbol{u} = 0, \tag{S11}$$

where $\boldsymbol{u}$ is the fluid velocity, $\eta$ is its viscosity, $p$ is the pressure and $\boldsymbol{f}^{\text{ext}}$ are external forces. For pointlike forces $\boldsymbol{f}^{\text{ext}} = f(\boldsymbol{r})\boldsymbol{e}\delta(\boldsymbol{r})$, where $\boldsymbol{e}$ is the predator's and prey's swimming direction, Eq. (S11) can be solved giving

$$\boldsymbol{u} = f(\boldsymbol{r})\mathcal{O}(\boldsymbol{r}), \tag{S12}$$

where

$$\mathcal{O}(\boldsymbol{r}) = \frac{1}{8\pi\eta r}(\mathbf{I} + \hat{\boldsymbol{r}} \otimes \hat{\boldsymbol{r}}), \tag{S13}$$

is the Oseen tensor, $\mathbf{I}$ the identity matrix, $\hat{\boldsymbol{r}} = \boldsymbol{r}/r$ the unit vector in $\boldsymbol{r}$ direction, $r = |\boldsymbol{r}|$ the length of $\boldsymbol{r}$, and $\otimes$ the tensor product.

## 4.1 Force monopole

The external force induced to the fluid by a force monopole including the contribution from our potential barrier is

$$\boldsymbol{f}_i^{\text{ext}} = \beta(v_i + \alpha_i x)\boldsymbol{e}, \tag{S14}$$

where $i = 1, 2$ gives the force from prey and predator respectively and $\beta$ is a constant that depends on the specific geometry and swimming mechanism of the microswimmer. We assume that our swimmers are constrained to only move in the $x$-direction, and their orientation is fixed to the $x$-direction, such that we can simplify $\boldsymbol{r} = x\boldsymbol{e}$. Plugging the force Eq. (S14) into Eq. (S12) and considering that the prey only experiences the flow of the predator and vice versa, gives

$$u_1 = \frac{\beta}{4\pi\eta}\left(\alpha_2 \text{sign}(x_1 - x_2) + \frac{v_2}{|x_1 - x_2|}\right), \tag{S15}$$

$$u_2 = \frac{\beta}{4\pi\eta}\left(\alpha_1 \text{sign}(x_2 - x_1) + \frac{v_1}{|x_1 - x_2|}\right), \tag{S16}$$

where $u_1$ is the flow experienced by the prey ($x_1$) and $u_2$ the flow experienced by the predator ($x_2$).

## 4.2 Force dipole

The flow field induced by a force dipole is given by

$$\boldsymbol{u}_i(\boldsymbol{r}) = \frac{\beta\lambda v_i}{8\pi\eta r^2}\left[3\left(\frac{\boldsymbol{r}}{r} \cdot \boldsymbol{e}\right)^2 - 1\right]\frac{\boldsymbol{r}}{r}, \tag{S17}$$

where $\alpha$ is a constant that depends on the distance between dipole and the specific swimming mechanism. We combine this with the force induced by the potential barrier which reads $\boldsymbol{f}_i^{\text{ext}} = \beta\alpha_i x\boldsymbol{e}$. Again, we assume that the motion of predator and prey is constrained to the $x$-direction. Summing the flow field induced by the force dipole and the potential barrier yields

$$u_1 = \frac{\beta \text{sign}(x_1 - x_2)}{4\pi\eta}\left(\alpha_2 + \frac{v_2\lambda}{(x_1 - x_2)^2}\right), \tag{S18}$$

$$u_2 = \frac{\beta \text{sign}(x_2 - x_1)}{4\pi\eta}\left(\alpha_1 + \frac{v_1\lambda}{(x_1 - x_2)^2}\right). \tag{S19}$$





\end{document}